\documentclass[10pt,twocolumn,showpacs,longbibliography,amsmath,amssymb,osajnl,floatfix,superscriptaddress]{revtex4-1}

\usepackage{graphicx}
\usepackage{dcolumn}
\usepackage{color}
\usepackage{txfonts}
\usepackage{microtype}
\usepackage{bm}

\begin{document}


\title{Spin-polarization effects of an ultrarelativistic electron beam in an ultraintense two-color laser pulse}


\author{Huai-Hang Song}
\affiliation{Beijing National Laboratory for Condensed Matter
Physics, Institute of Physics, CAS, Beijing 100190, China}
\affiliation{School of Physical Sciences, University of Chinese
Academy of Sciences, Beijing 100049, China}

\author{Wei-Min Wang}
\email{weiminwang1@ruc.edu.cn} \affiliation{Department of Physics,
Renmin University of China, Beijing 100872, China}
\affiliation{Beijing National Laboratory for Condensed Matter
Physics, Institute of Physics, CAS, Beijing 100190, China}

\author{Jian-Xing Li}
\email{jianxing@xjtu.edu.cn} \affiliation{MOE Key Laboratory for
Nonequilibrium Synthesis and Modulation of Condensed Matter, School
of Science, Xi'an Jiaotong University, Xi'an 710049, China}

\author{Yan-Fei Li}
\affiliation{MOE Key Laboratory for Nonequilibrium Synthesis and
Modulation of Condensed Matter, School of Science, Xi'an Jiaotong
University, Xi'an 710049, China}

\author{Yu-Tong Li}
\affiliation{Beijing National Laboratory for Condensed Matter
Physics, Institute of Physics, CAS, Beijing 100190, China}
\affiliation{School of Physical Sciences, University of Chinese
Academy of Sciences, Beijing 100049, China}
\affiliation{Songshan Lake Materials Laboratory, Dongguan,
Guangdong 523808, China}

%

\date{\today}

\begin{abstract}

Spin-polarization effects of an ultrarelativistic electron beam
head-on colliding with an ultraintense two-color laser pulse are
investigated comprehensively in the quantum radiation-dominated
regime. We employ a Monte Carlo method, derived from the recent work
of [Phys. Rev. Lett. {\bf 122}, 154801 (2019)], to calculate the
spin-resolved electron dynamics and photon emissions in the local
constant field approximation. We find that electron radiation
probabilities in adjacent half cycles of a two-color laser field are
substantially asymmetric due to the asymmetric field strengths, and
consequently, after interaction the electron beam can obtain a total
polarization of about 11\% and a partial polarization of up to about
63\% because of radiative spin effects, with currently achievable
laser facilities, which may be utilized in high-energy physics and
nuclear physics. Moreover, the considered effects are shown to be
crucially determined by the relative phase of the two-color laser
field and robust with respect to other laser and electron beam
parameters.

\end{abstract}


\maketitle


\section{Introduction}

As one of the intrinsic properties carried by electrons, the spin
has been extensively studied and utilized in the high-energy physics
\cite{Moortgat2008pr,mane2005rpp,abbott2016prl}, materials science
\cite{zutic2004rmp}, and plasma physics
\cite{marklund2007prl,bordin2007njp}. As known, the relativistic
polarized electrons are commonly generated via two methods. The
first extracts polarized electrons from a photocathode
\cite{Pierce_1976} or spin filters \cite{Batelann_1999,Dellweg_2017,
Dellweg_2017PRA}, and then employs a conventional accelerator or a
laser wakefield accelerator \cite{Wen_2018} to accelerate them into
the relativistic realm. The second directly polarizes a relativistic
electron beam in a storage ring via using the radiative polarization
effect (Sokolov-Ternov effect)
\cite{Sokolov_1964,Sokolov_1968,Baier_1967,Baier_1972,
Derbenev_1973}. However, the latter typically requires a long
polarization time of about minutes$\sim$hours because of the low
static magnetic field at the Tesla scale.

Recently, the rapid development of ultrashort (duration $\sim$ tens of
femtoseconds) ultraintense (peak intensity $\sim 10^{22}~ \rm
W~cm^{-2}$, and the corresponding magnetic field  $\sim 4\times
10^5$ Tesla) laser techniques \cite{eli,xcels} is providing
opportunities to investigate electron polarization effects in such
strong laser fields, analogous to the Sokolov-Ternov effect. A
plenty of theoretical works have been performed in nonlinear Compton
scattering, e.g., see
\cite{Panek_2002,Kotkin_2003,Karlovets_2011,Boca_2012,
Krajewska_2013} and the references therein. However, only a small
polarization can be obtained in a monochromatic laser field
\cite{Ivanov_2004} or a laser pulse \cite{Seipt_2018}. A  setup of
strong rotating  electric fields \cite{Sorbo_2017,Sorbo_2018} shows
a rather high polarization, when the electrons are trapped at the
antinodes of the electric field. Unfortunately,  this case may only
occur for linearly polarized laser pulses of intensities $\gtrsim
10^{26}~ \rm W~cm^{-2}$ \cite{Gonoskov_2014}, which is much beyond
current achievable laser intensities. Recently, a scheme with an
elliptically polarized laser pulse has been proposed to split the
electrons with different spin polarizations through spin-dependent
radiation reaction \cite{li2019prl}, and consequently, to reach a
polarization above 70\%. Also, a similar setup can be used to
generate a positron beam with a polarization up
to 90\% due to asymmetric spin-dependent pair production
probabilities \cite{Wan_2019}.

Previous works indicate that the total polarization of all electrons
in monochromatic laser pulses are negligible because of the
symmetric laser field. In other words, asymmetric laser fields may
result in a considerable  polarization. The well-known asymmetric  two-color
laser configuration has been widely adopted in generation of
Terahertz radiation
\cite{Kim2007oe,Andreeva2016prl,Zhang2018prl,Wang2017pra}, high
harmonic wave generation \cite{Dudovich2006np,Chen2014pnas}, and
laser wakefield acceleration \cite{Zeng2015prl}. Recently, it is
also proposed to generate polarized positron beams through
multiphoton Breit-Wheeler pair production  \cite{Chen_2019}.
However, employing such two-color laser configuration to directly 
polarize the ultrarelativistic electron beam via nonlinear Compton scattering  is still an open challenge.

\begin{figure}[t]
\centering
\includegraphics[width=1.0\linewidth]{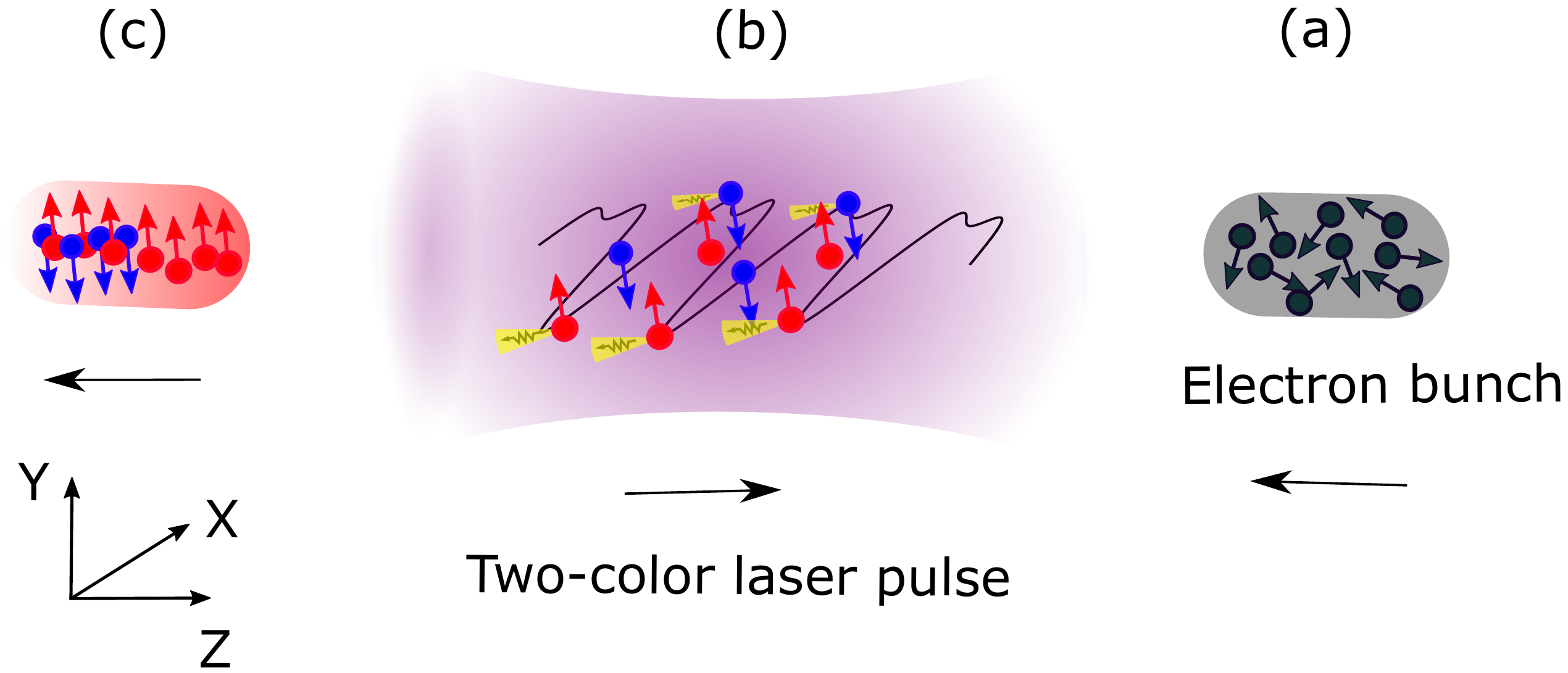}
\caption{\label{fig1} The interaction scenario of an
ultrarelativistic electron beam head-on colliding with  an
ultraintense two-color laser pulse. (a) An unpolarized  electron
beam propagates along the $-z$ direction, which can be obtained from
a laser wakefield accelerator. (b) The interaction between the
electron beam and the two-color laser pulse, polarizing along $x$ axis and propagating along $+z$ direction, results in photon
emissions and spin-flip transitions of the electrons. (c) A transversely polarized (in $y$ axis)
electron beam  can be achieved after interaction. The black-random [in (a)], red-up, and blue-down arrows [in (b) and (c)] indicate the unpolarized, spin-up,
and spin-down electrons with respect to $+y$ direction, respectively. The violet curve and the yellow signs with black lines in (b) indicate
the two-color laser field and emitted photons, respectively. }
\end{figure}

In this work, the polarization effects of an ultrarelativistic
electron beam head-on colliding with a currently achievable
ultraintense two-color laser pulse are comprehensively investigated
in quantum radiation-dominated regime (see the interaction scenario
in Fig.~\ref{fig1}). During the interaction, the radiation
probabilities of electrons in the positive and negative half cycles
of the two-color laser field are substantially asymmetric. Thus,
after interaction considerable total polarization and partial
polarization can be obtained. We find that the
relative phase $\phi$ of the two-color laser pulse is crucial to
determine the polarization effects. In particular, when
$\phi=\pi/2$, the laser field strengths in negative half cycles are
much higher than those in the positive cycles, and consequently, more
photons of higher energies are emitted in the negative half cycles.
Accordingly, the electron spins more probably flip to the direction
antiparallel to the laser magnetic field in the electron's rest
frame, assumed to be the instantaneous spin quantization axis (SQA)
\cite{li2019prl}, and those electrons have lower remaining energies
due to radiation-reaction effects \cite{Piazza2012}. As $\phi$
changes, the considered effects are weakened until complete
disappearance in the case of $\phi=0$. Moreover, the impacts of the
laser and electron beam parameters on the considered effects are
studied, and optimal parameters are analyzed.

This paper is organized as follows. Section~\ref{model} presents the employed Monte Carlo simulation model. In Sec.~\ref{results},
the polarization effects of the ultrarelativistic electron beam in the two-color laser pulse are shown and analyzed,  and  the
impacts of the laser and electron beam parameters on the
polarization effects are also investigated. Finally, a brief summary is
given in Sec.~\ref{conclusion}.

\section{the theoretical model}
\label{model}

The quantum electrodynamics (QED) effects in the strong field are
governed by the dimensionless and invariant QED parameter
$\chi\equiv(e\hbar/m^3c^4)\sqrt{-|F_{\mu\nu}p^{\nu}|}$ \cite{Ritus_1985}, where
$F_{\mu\nu}$ is the field tensor, $p^{\nu}$ the electron’s
4-momentum, and the constants $\hbar$, $m$, $e$ and $c$ are the reduced
Planck constant, the electron mass and charge, and the velocity of
light, respectively. The normalized laser field amplitude parameter $\xi\equiv
eE_0/(mc\omega_L)\gg1$ and the QED parameter $\chi \lesssim1$ are
considered to ensure that the coherence length of the photon emission is
much smaller than the laser wavelength \cite{Ritus_1985}. Here $E_0$ and $\omega_L$ are the laser
field amplitude and angular frequency, respectively. The
spin-dependent probability of photon emission
in the local constant field approximation can be written (summed up by photon
polarization and electron spin after photon emission) as
\cite{li2019prl, Baier_1973}

\begin{eqnarray}\label{eq1}
\frac{d^2\overline W_{rad}}{dudt}&=&\frac{\alpha m^2c^4}{\sqrt{3}\pi \hbar \varepsilon_e}\left[\left(1-u+\frac{1}{1-u}\right)K_{2/3}(y)\right.\nonumber\\
&&\left.-\int_{y}^{\infty}K_{1/3}(x)dx -(\bm S_i \cdot \bm \zeta)uK_{1/3}(y) \right],
\end{eqnarray}
where $K_{\nu}$ is the modified Bessel function of the order of
$\nu$, $y=2u /[3(1-u)\chi]$, $u=\varepsilon_\gamma / \varepsilon_e$,
$\varepsilon_e$ the electron energy before radiation,
$\varepsilon_\gamma$ the emitted photon energy, and $\alpha$ the
fine structure constant. The last term in Eq.~(\ref{eq1}) is a
spin-dependent addition, where $\bm S_i$ is the initial spin
 vector of an electron before photon emission, and $\bm \zeta=\bm \beta\times \bm \hat{a}$. $\bm \beta$ is the electron velocity normalized by $c$, and $\bm \hat{a}=\bm a/|\bm a|$ is the electron acceleration. By averaging
over the initial electron spin $\bm S_i$, the widely employed spin-free radiation
probability can be obtained \cite{Sokolov2010,Elkina2011,Ridgers_2014,Green2015,Harvey_2015}.
The spin vector
$\bm S$ = $(S_x, S_y, S_z)$, and $|\bm S|=1$.

The stochastic photon emission by an electron can be calculated via using
the conventional QED Monte-Carlo algorithm \cite{Ridgers_2014}
with a spin-dependent radiation probability given by
Eq.~(\ref{eq1}). The electron dynamics in the external laser field
is described by classical Newton-Lorentz equations, and its spin
dynamics is calculated according to the
Thomas-Bargmann-Michel-Telegdi equation \cite{Thomas_1926,Thomas_1927,Bargmann_1959, Walser_2002}.
After photon emission, the electron spin is assumed to flip either parallel
or antiparallel to the instantaneous SQA (along $\bm \zeta$) with a probability given in Ref.~\cite{li2019prl}.
Note that, as shown in the last term of Eq.~(\ref{eq1}), when the spin vector $\bm S_i$ is
antiparallel to the instantaneous SQA, the electron has a higher
probability to emit a photon.

\section{Results and analysis}
\label{results}

\subsection{\label{rr}Simulation setup}

\begin{figure*}[t]
\centering
 \includegraphics[width=1.0\linewidth]{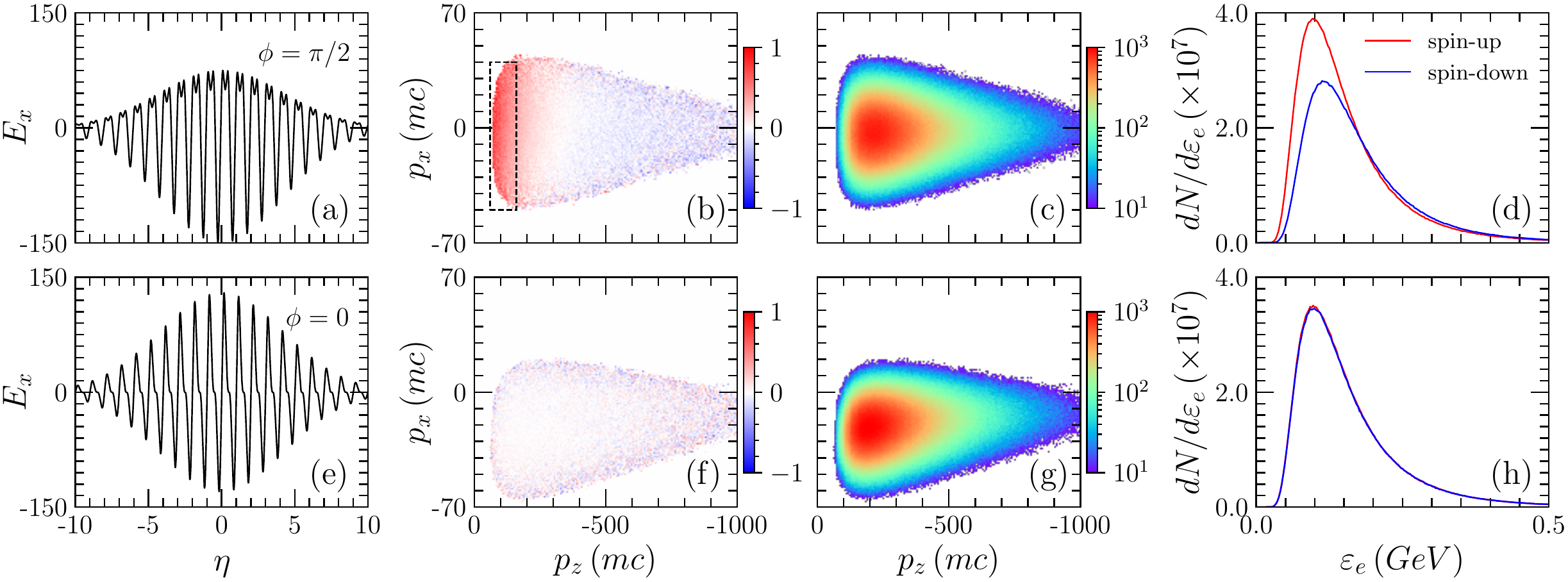}
\caption{\label{fig2} [(a), (e)] The laser field $E_x$ with respect to $\eta$. [(b), (f)]
Distribution of the average polarization
$\overline{S}_y$ versus longitudinal and transverse momenta $p_z$ and
$p_x$, respectively. [(c), (g)] Number density distributions of
electrons versus $p_z$ and $p_x$. [(d), (h)] Energy spectra of
spin-up and spin-down electrons, respectively. Note that ``spin-up'' and ``spin-down'' indicate 
the electron spin parallel and antiparallel to the $+y$ axis, respectively.  Upper panels (a)-(d)
indicate the simulation results with
$\phi=\pi/2$, and lower panels (e)-(h) with $\phi=0$.}
\end{figure*}

In our simulations, the fundamental laser pulse of a wavelength
$\lambda_0=1.0~\rm \mu m$ and the second harmonic pulse have the
same duration, transverse profile, and linear polarization along the
$x$ direction. They propagate along the $+z$ direction and their
combined electric field can be expressed as $E_x
\propto[\xi_1sin(\omega_L\eta)+\xi_2sin(2\omega_L\eta+\phi)]$, where
$\xi_1$ and $\xi_2$ are the normalized amplitudes of the fundamental
and the second-harmonic pulses, respectively, $\eta=(t-z/c)$, and
$\phi$ is the relative phase. We employ a three-dimensional
description of the tightly-focused laser pulse with a Gaussian
temporal profile with the fifth order $(\sigma_0/z_r)^5$ in the
diffraction angle \cite{Salamin2002prl}, where $z_r=k_L\sigma_0^2/2$
is the Rayleigh length, $k_L=\omega_L/c$ the wave vector, and
$\sigma_0$ the waist radius.

In our first simulation, we take the laser peak amplitude
$\xi_1=2\xi_2=100$ (corresponding to  the peak intensity $I_1=4I_2
=1.37 \times 10^{22}~\rm W~cm^{-2}$), and full width at half maximum
(FWHM) duration $\tau_0$ = 10 $T_0$ (33~fs), where $T_0$ is the
laser period. Considering that the different Rayleigh lengths of
two-color laser pulses, we firstly take the waist radius as infinity
for simplicity, and then we will discuss the finite waist effects.
Our simulations will show that the results in the plane wave case
are very close to the ones with $\sigma_0 \geq 5~\mu m$. An
unpolarized cylinderical electron beam is employed, including $10^7$ electrons
with initial mean energy $\varepsilon_0$ = 1.5 GeV (corresponding to
the relativistic factor $\gamma_0\approx2935$), energy spread
$\Delta \varepsilon_0 / \varepsilon_0=10\%$, transversely  Gaussian
profile with a radius $r_1=3~\rm \mu m$, and longitudinally uniform
profile with a length $r_2=5~\rm \mu m$. This kind of electron bunch
can be obtained by laser wakefield accelerators
\cite{Leemans2014,Leemans_2019}

During the head-on collision, one could assume the momenta of
ultrarelativistic electrons to be approximately along the initial
moving direction, i.e.,  the $-z$ direction, due to $\gamma_0 \gg
\xi_1$. Hence, the magnetic fields experienced by the electrons in
their rest frames are along the $y$ axis. Note that ``spin-up'' and ``spin-down'' indicate the electron spin parallel and antiparallel to the $+y$ axis, respectively.

\subsection{\label{rr} Electron polarization via radiative spin effects}

\begin{figure}[t]
\centering
\includegraphics[width=1.0\linewidth]{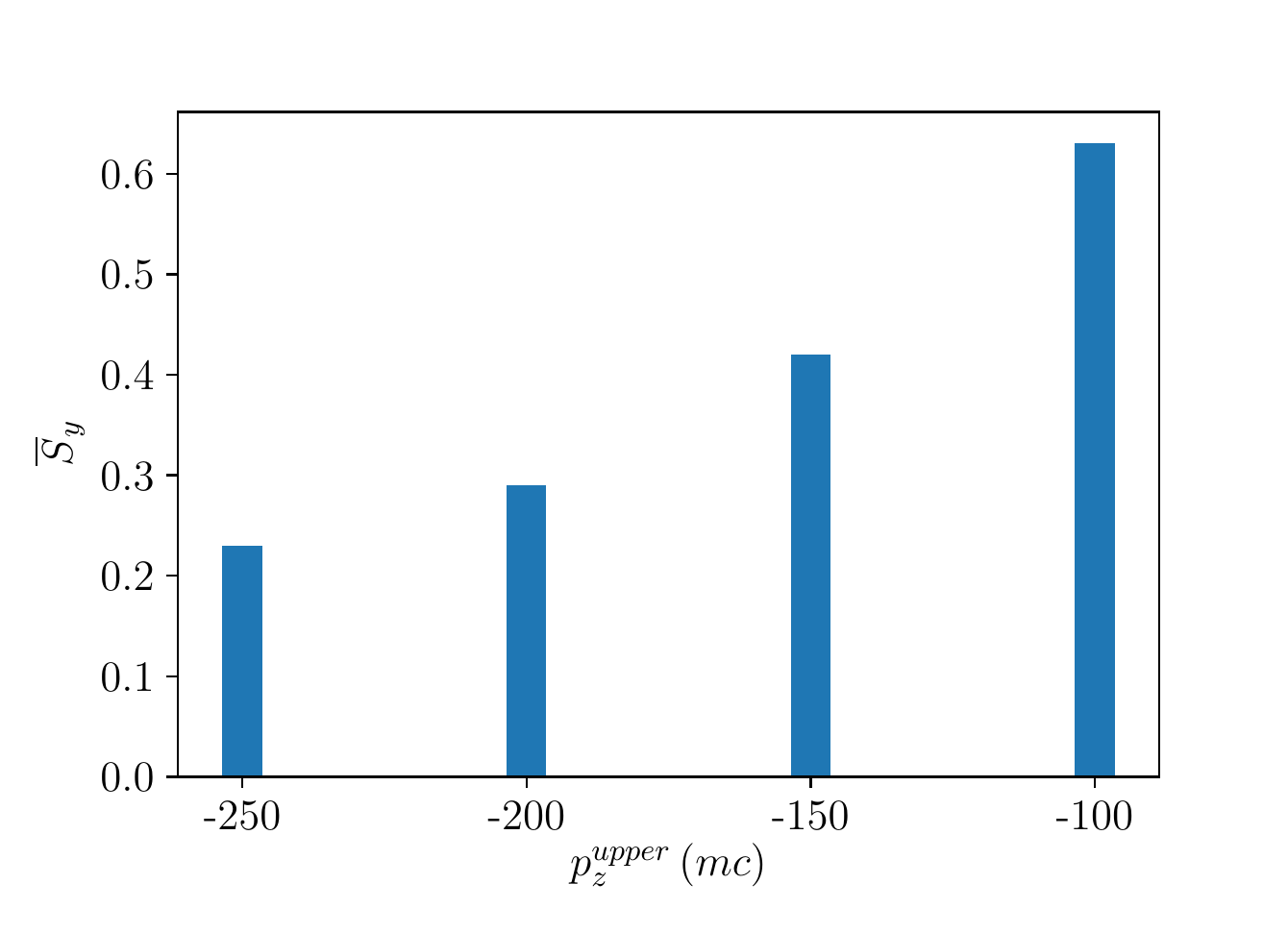}
\caption{\label{fig3} The average polarization  $\overline{S}_y$ of the
electrons with different cutoff $p_z$.}
\end{figure}

The combined electric field of the two-color laser pulse has a highly
asymmetric envelope profiles in the positive and negative half
cycles when $\phi=\pi/2$, as shown in Fig.~\ref{fig2}(a). The
electrons in the negative half cycles with higher field strengths
have a larger QED parameter $\chi$, which causes more photons with
higher energies to be emitted than those in the positive half
cycles. In the negative half cycles, the instantaneous SQA (along $\bm \zeta=\bm \beta\times \bm \hat{a}$) is
along $-y$ direction, therefore, after photon emission the electron spin is more probably antiparallel to the SQA, i.e., $+y$ direction \cite{li2019prl}. This
results in generation of more spin-up (with respect to $+y$ direction) electrons, as shown in
Fig.~\ref{fig2}(d). Accordingly, the total polarization
of the whole electron beam is about $11\%$.
Moreover, due to radiation-reaction effects, more spin-up electrons have lower energies [see Fig.~\ref{fig2}(b)]. In
the region of  $|p_z|<160~mc$ marked  by the black dotted box, the polarization of $14\%$ electrons is
 above $40\%$. Further, if one
filters high-energy electrons, the polarization of remaining electrons with
$|p_z|<100~mc$ is up to about $63\%$, as shown in
Fig.~\ref{fig3}. Obviously, the energy-dependent polarization could
provide a way to generate a highly-polarized electron beam by
choosing electron energy. And, it may
present an experimental scheme to verify the theory of the
spin-dependent radiation reaction. Note that the polarization of laser-driven ultrarelativistic electron beams
can be measured via the polarimetry of nonlinear Compton scattering \cite{li2019arxiv}.

As $\phi=0$, the combined electric field has symmetric envelope
profiles in the positive and negative half cycles, as shown in
Fig.~\ref{fig2}(e).  Such a laser field cannot generate more spin-up
or spin-down electrons via nonlinear Compton scattering, as observed
in Fig.~\ref{fig2}(h), because  the polarization of electrons induced in
the positive and negative cycles counteracts each other. One can
notice in Figs.~\ref{fig2}(f) and (g) that the electrons can acquire
a non-zero drift velocity in a such field configuration due to
asymmetry in the laser vector potential
\cite{Zhang2018prl,Wang2017pra} and radiation reaction
\cite{Tamburini2014pre}. Besides, it is shown in  Figs.~\ref{fig2}(d)
and (h) that the energy spectra of the spin-up and
spin-down electrons both become broader compared with the initial
quasi-monoenergetic spectrum, because the electrons lose energies via
stochastic photon emissions.\\

\begin{figure}[t]
\centering
\includegraphics[width=1.0\linewidth]{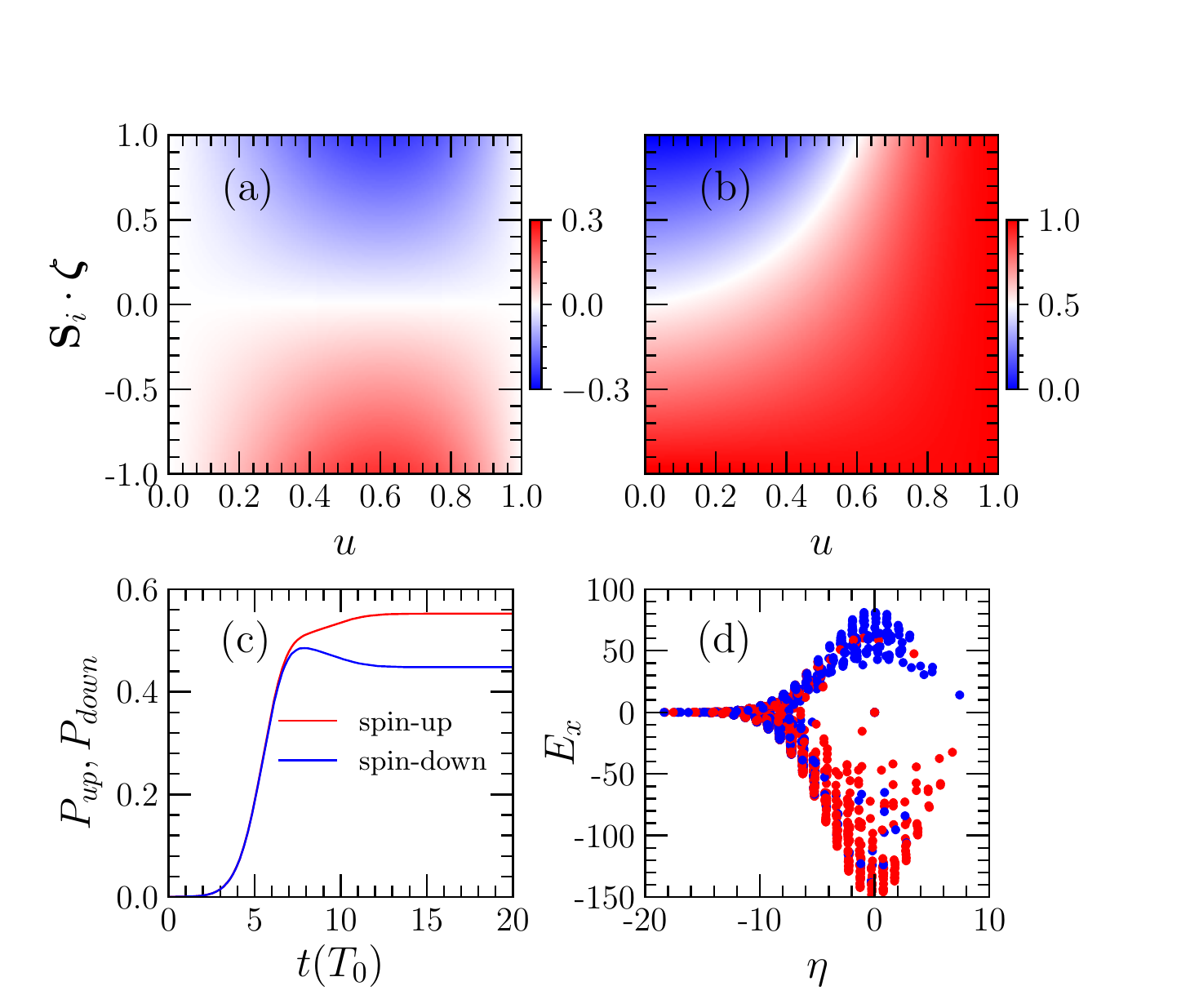}
\caption{\label{fig4} (a) The ratio of the last term in
Eq.~(\ref{eq1}) to the first two terms versus $\bm S_i \cdot \bm
\zeta$ and $u$, where the QED parameter $\chi \approx 1.1$. (b) The probability that an electron spin
flips to the direction antiparallel to the instantaneous SQA after emitting a photon, with $\chi\approx1.1$.
(c) Evolution of the proportions of the spin-up ($P_{up}$, red line) and spin-down
($P_{down}$, blue line) electrons, respectively. (d) Tracks of
spin-flip dynamics of 2000 electrons chosen randomly, where the red
and blue points indicate electrons flipping to spin-up and
spin-down, respectively, after photon emissions. Note that ``spin-up'' and ``spin-down'' indicate
the electron spin parallel and antiparallel to the $+y$ axis, respectively. Other laser and electron parameters are the same as those in Fig.~\ref{fig2}. }
\end{figure}

To analyze the reasons of the polarization effects, Fig.~\ref{fig4} shows the details of the evolution of the electron
spin flips in the two-color laser field with $\phi=\pi/2$. When
interacting with the laser field, electrons emit photons, and
the spin flips  either parallel or antiparallel
to the instantaneous SQA \cite{li2019prl}. The formed electron
polarization can significantly affect the photon emission
according to the last term in Eq.~(\ref{eq1}). With $\bm S_i \cdot
\bm \zeta=-1$, i.e., the electron spin is antiparallel to the instantaneous SQA, the emission probability could be enhanced by about $30\%$,
oppositely, it could be decayed by about $30\%$ with $\bm S_i \cdot \bm
\zeta=1$, as shown in Fig.~\ref{fig4}(a).

In Fig.~\ref{fig4}(b), we demonstrate the probability that an electron spin
flips to the direction antiparallel to the instantaneous SQA
after emitting a photon. One can see that the spin-flip probability depends on
both the electron spin direction and the emitted photon energy.  With
$\bm S_i \cdot \bm \zeta<0$, the electron
spin very likely flips even though the emitted photon has a low
energy. With $\bm S_i \cdot \bm \zeta>0$, the spin flip arises with
a high probability when the emitted photon energy is high enough.
Basically, the electron spin tends to flip to the direction antiparallel to
the SQA. Note that above analysis holds at high laser
intensities [$\chi\approx1.1$ is employed in Figs.~\ref{fig4}(a)
and (b)]. When the laser intensity is low and the resulting QED
parameter $\chi\sim\xi$ is also small, the photon energy is usually much
lower than that of electron, $u=\varepsilon_\gamma/\varepsilon_e \sim \chi$. Hence,
contributions of the electron spin term to the spin-flip
probability  as well as the radiation probability
given by Eq. (1)  can be ignored.

In Fig.~\ref{fig4}(c), we show the ratios of the spin-up
and spin-down electron numbers to the total electron number, respectively.
When the electron beam collides with the rising edge of the laser
pulse at $t\lesssim 7~T_0$, the electrons gradually flip to spin-up or
spin-down with nearly the same probability, due to the low laser
field strength and small $\chi$. As the electrons approach the laser
pulse peak around $t\approx10~T_0$, $\chi$ grows to about 1.1, and more
spin-up electrons are generated accompanied with higher energy
emitted photons. The similar results can be found in
Fig.~\ref{fig4}(d), in which we randomly choose 2000 electrons and
track their dynamics. It is clearly shown that in the strong laser
field region the spin flips are significant. In the negative half
cycles of the electric field, the instantaneous SQA is along $-y$ direction, and the electrons incline to flip to
spin-up, i.e., $+y$ direction. Oppositely, they tend to flip to spin-down, i.e., $-y$ direction, in the positive
half cycles. Because the field strengths in the negative half cycles
are stronger, more electrons probably flip to spin-up, and consequently, a
polarized electron beam is obtained.

\subsection{\label{rr}Impacts of the laser and electron beam parameters on the total polarization of the electron beam}

\begin{figure}[t]
\centering
\includegraphics[width=1.0\linewidth]{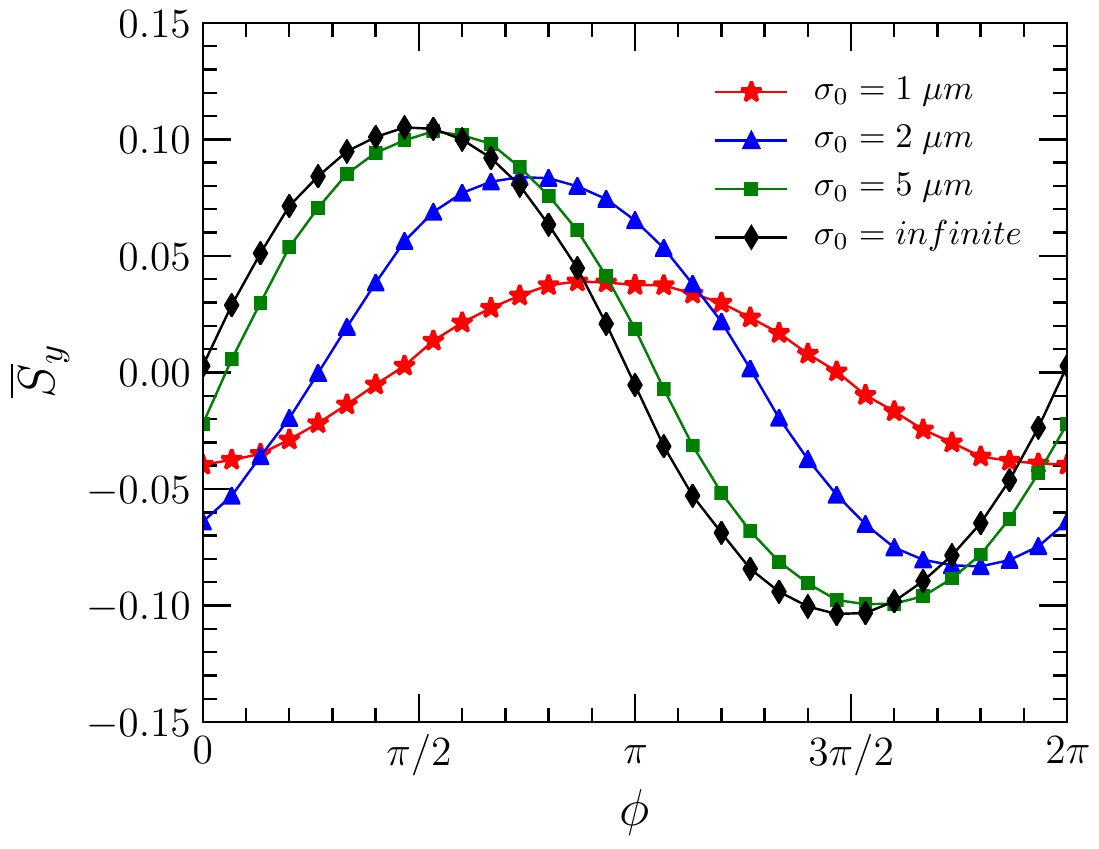}
\caption{\label{fig5} The average polarization  $\overline{S}_y$ as a function of  the relative
phase $\phi$ of the two-color laser pulse with
different waist radii. Other laser and electron parameters are the same as those in Fig.~\ref{fig2}.}
\end{figure}

We further study the impacts of the laser and electron beam parameters on the total polarization of the electron beam. In Fig.~\ref{fig5},
we change the relative phase $\phi$ with different waist radius
$\sigma_0$. When $\sigma_0$ approaches infinite, i.e., the plane
wave case, shown by the black curve with diamonds, the total
polarization is zero at $\phi=0$, increases gradually to the maximum
at $\phi=\pi/2$, and then decreases to zero at around $\phi=\pi$.
Within the range of $\phi$ between $\pi$ and $2\pi$, the same result
can be observed except that the polarization turns negative,
i.e., more spin-down electrons are generated. This is because the
laser strengths in the negative half cycles are higher with $\phi\in
(0,~\pi)$, while the ones in the positive half cycles are higher with
$\phi\in (\pi,~2\pi)$. The dependency of the
polarization on $\phi$ roughly follows the character of the function $\sin{(\phi)}$, similar to the
THz generation dependency on $\phi$ \cite{Kim2007oe}, which results from the
dependency of laser pulse envelope asymmetry between the positive
and the negative half cycles on $\phi$.

When we take the laser waist radius as $\sigma_0=5~\rm \mu m$, the
dependency of the polarization on $\phi$ is still close to the plane
wave case. However, as the waist radius is further decreased to
$2~\rm \mu m$ and $1~\rm \mu m$, the dependency deviates gradually
from the plane wave case. The maximum of the polarization does not
appear at $\phi=\pi/2$ and $\phi=3\pi/2$, and the maximum is reduced
significantly. These characters can be explained by the
different Rayleigh lengths between the fundamental laser pulse and
the second-harmonic one. As the pulses propagate, the envelope of
the combined laser field as well as the the ratio of
two laser amplitudes walk off. They can remain the same as the
plane wave case only at the laser envelope peak. Therefore, the
asymmetry of the laser field with $\phi=\pi/2$ is weakened with the
decrease of the waist radius. To obtain a considerable polarization, the
laser waist radius should be taken as $\sigma_0 \gtrsim 5~\rm \mu m$.

\begin{figure}[t]
\centering
\includegraphics[width=1.0\linewidth]{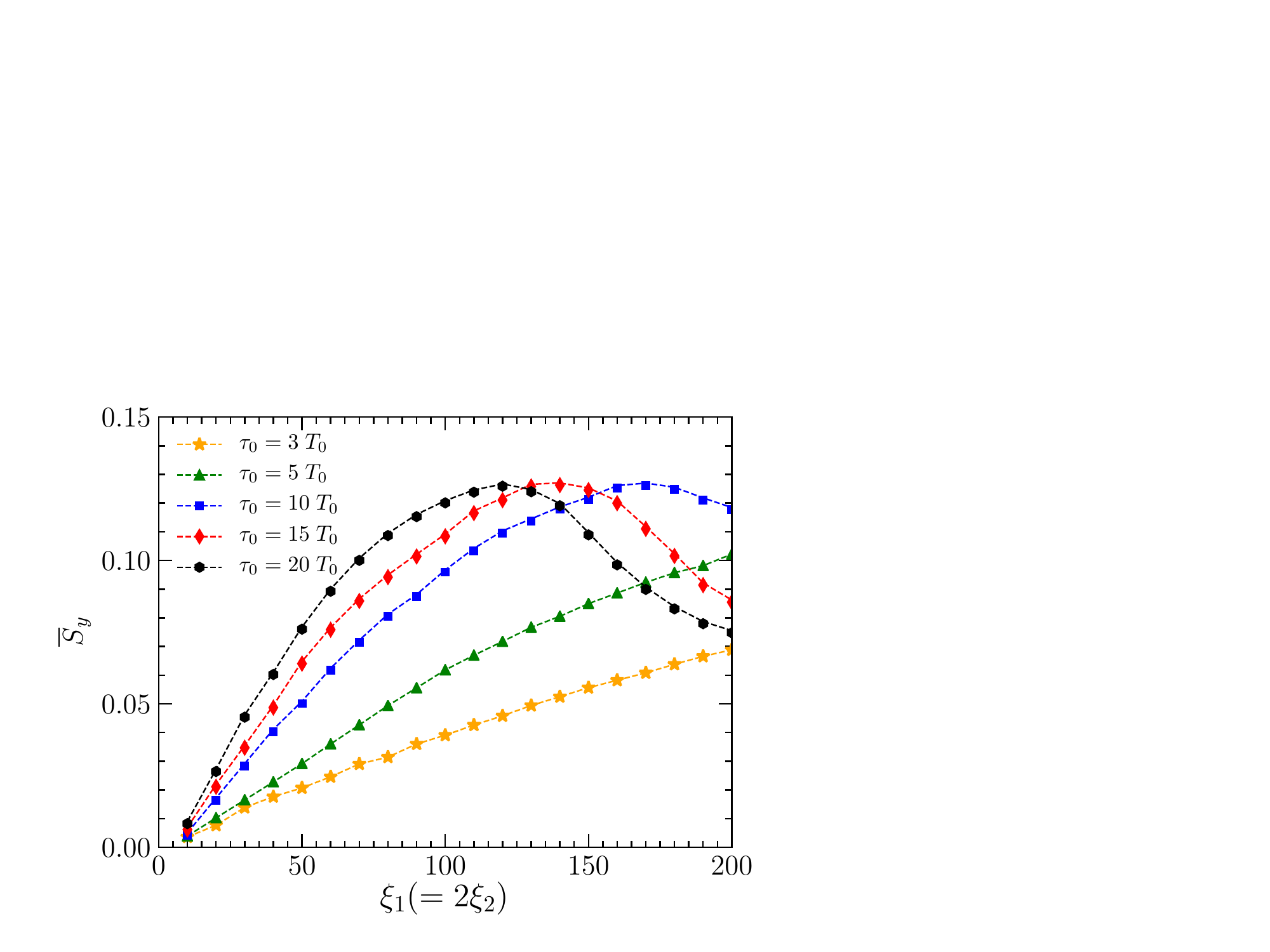}
\caption{\label{fig6} The average polarization  $\overline{S}_y$
as a function of $\xi_1$ with different pulse
durations, where the initial electron energy
$\varepsilon_0=1.5$ GeV. Here, $\sigma_0=5~ \rm \mu m$ is taken.}
\end{figure}

\begin{figure}[t]
\centering
\includegraphics[width=1.0\linewidth]{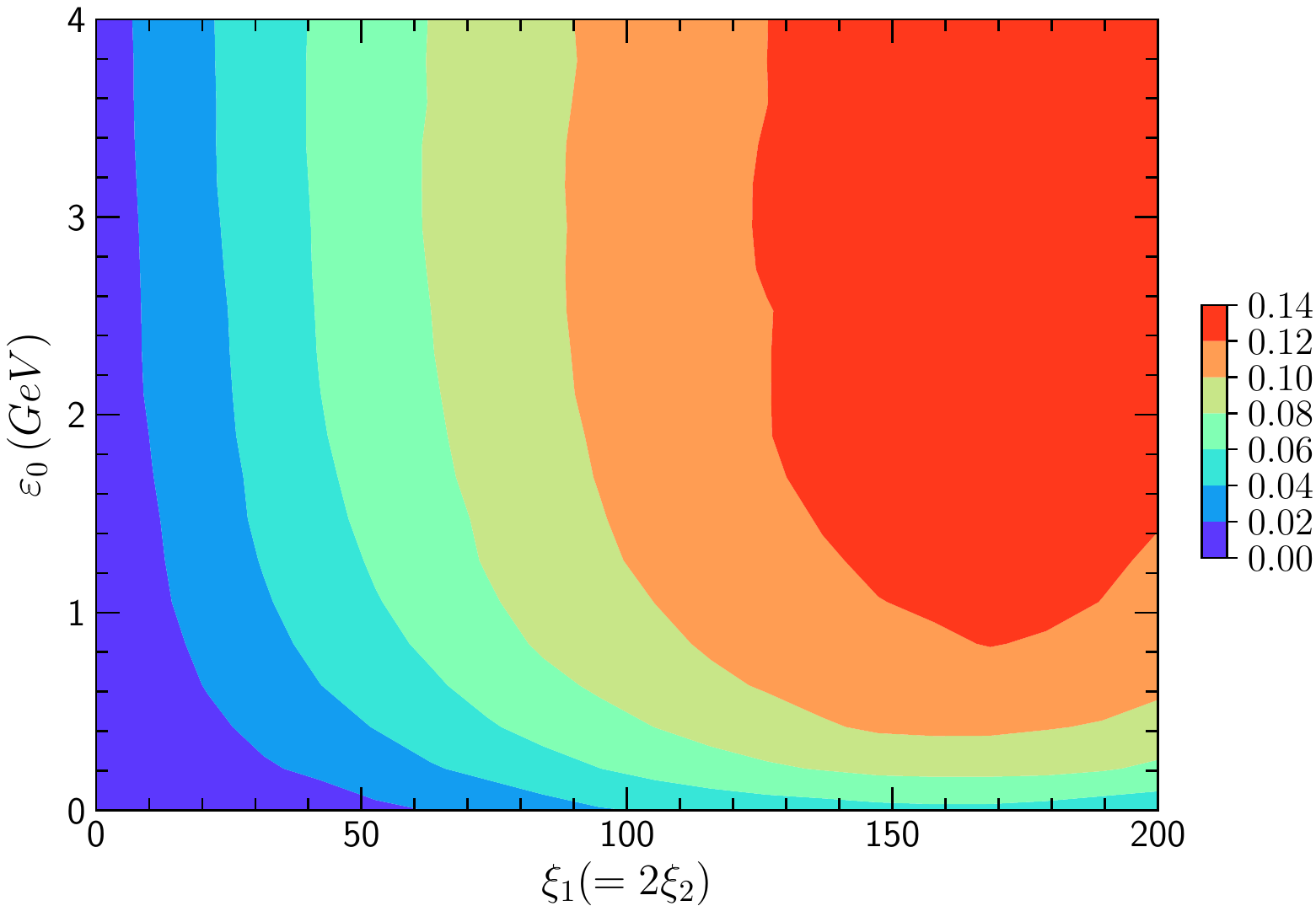}
\caption{\label{fig7} The average polarization $\overline{S}_y$
versus the laser peak intensity and the initial electron
energy $\varepsilon_0$. Here, $\tau_0=10~ \rm T_0$ and $\sigma_0=5 ~\rm \mu m$ are taken.}
\end{figure}

Furthermore, we investigate the impacts of the laser peak intensity
and pulse duration on the considered effects, as presented in
Fig.~\ref{fig6}. We employ $\phi=\pi/2$, $\sigma_0=5~\rm \mu m$, and
$\xi_1=2\xi_2$.  When the laser duration $\tau _0=10~T_0$ (FWHM
$\sim$ 33 fs), with enhancing $\xi_1$ (as well as $\xi_2$) the polarization first
increases, and then decreases. The similar results are also observed
with longer durations, e.g., $\tau _0=15~T_0$ and $20~T_0$. However,
the peak appears at a lower $\xi_1$ for a longer duration. As the
duration is decreased to  $\tau _0=5~T_0$ and $3~T_0$, only a
monotonical increase appears within the $\xi_1$ region considered.
It is expected that the polarization will decay if
higher $\xi$ is adopted. One can also observe that in the increasing
region the polarization is higher for a longer duration when the laser amplitude $\xi_1$ is fixed. 
The polarization first grows with both of the
laser pulse duration and amplitude because of the probabilities of photon
emission and electron spin flip $\sim\chi\tau_0\sim\xi\tau_0$. Due
to photon emission, the electrons lose their energies. Provided the
laser pulse duration is too long, the electrons could lose their
main energies in the rising edge of the laser pulses, and the
effective laser fields experienced by the electrons are much lower
than that at the laser pulse peak. This could causes that the
polarization decays with the increase of $\xi_1$.

Finally, we study the combined role of the initial electron energy
$\varepsilon_0$ and the laser peak amplitude, as shown in Fig.~\ref{fig7}. It is found that a high laser amplitude (e.g., $\xi_1\gtrsim 100$) is
necessary to obtain a high total polarization. With a high laser amplitude, the electron beam
energy could be flexible in a large range from hundreds of MeV to
few GeV. On the other hand, even though a high electron beam energy is taken (e.g., $\varepsilon_0\approx$ 4 GeV),
the total polarization is relative low.

\section{conclusion}
In summary, we have investigated the spin polarization effects of an
ultrarelativistic electron beam head-on colliding with an
ultraintense two-color laser pulse. The asymmetry of the laser field 
in the processes of the photon emission and the electron spin-flip
transition causes  considerable total and partial
polarization. The polarization strongly depends on the relative
phase $\phi$ of the two-color laser pulse. When $\phi=\pi/2$, the
degree of a certain polarization reaches its peak. As $\phi$ is
taken as $3\pi/2$, the same degree is achieved, however, the
polarization turns opposite. Moreover, the spin-dependent radiation
reaction results in the high polarization of relative-low-energy
electrons, which provides a way to generate a highly polarized
electron beam by choosing electron energy, and may serve as a
signature of the spin-dependent radiation reaction in the QED
regime. \label{conclusion}

\section{acknowledgments}
This work was supported by National Key R\&D Program of China (Grant
No. 2018YFA0404801), National Natural Science Foundation of China
(Grants Nos. 11775302, 11874295 and 11804269), and Science Challenge
Project of China (Grant No. TZ2016005 and TZ2018005).

\bibliography{QEDspin}

\end{document}